\documentclass[prd,twocolumn,showpacs,amsmath,amssymb]{revtex4}
\usepackage{mathrsfs}
\usepackage{graphicx}
\usepackage[all]{xy}
\begin{document}
\title{Exact solutions of embedding the four-dimensional
perfect fluid in a five- or higher-dimensional Einstein spacetime
and the cosmological interpretations}
\author{Jie Ren$^{1,4}$}
\email{jrenphysics@hotmail.com}
\author{Xin-He Meng$^{2,3,4}$}
\author{Liu Zhao$^{2,4}$}
\affiliation{$^1$Theoretical Physics Division, Chern Institute of
Mathematics, Nankai University, Tianjin 300071, China}
\affiliation{$^2$Department of Physics, Nankai University, Tianjin
300071, China} \affiliation{$^3$BK21 Division of Advanced Research
and Education in physics, Hanyang University, Seoul 133-791, Korea}
\affiliation{$^4$Kavli Institute for Theoretical Physics China at
the Chinese Academy of Sciences (KITPC-CAS), Beijing, 100080, China}
\date{\today}
\begin{abstract}
We investigate an exact solution that describes the embedding of the
four-dimensional (4D) perfect fluid in a five-dimensional (5D)
Einstein spacetime. The effective metric of the 4D perfect fluid as
a hypersurface with induced matter is equivalent to the
Robertson-Walker metric of cosmology. This general solution shows
interconnections among many 5D solutions, such as the solution in
the braneworld scenario and the topological black hole with
cosmological constant. If the 5D cosmological constant is positive,
the metric periodically depends on the extra dimension. Thus we can
compactify the extra dimension on $S^1$ and study the
phenomenological issues. We also generalize the metric ansatz to the
higher-dimensional case, in which the 4D part of the Einstein
equations can be reduced to a linear equation.
\end{abstract}
 \maketitle

\section{Introduction}
Exact solutions play an important role in gravitational physics and
cosmology. The definition of Einstein spacetime is that the Einstein
tensor is zero or proportional to its metric tensor, i.e.,
$G_{\mu\nu}=-\Lambda g_{\mu\nu}$, where $\Lambda$ is the
cosmological constant. Our Universe is not an Einstein spacetime
because it contains matter. In the standard
Friedmann-Robertson-Walker (FRW) framework, the content of our
Universe is assumed to be perfect fluid in consistent with the
cosmological principle. Therefore, we are interested in exact
solutions that can describe the embedding of the perfect fluid in a
higher-dimensional Einstein spacetime, as a generalization of the
embedding only between Einstein spacetimes \cite{warp,zhao}. The
Cambell-Magaard theorem and its generalized versions \cite{cm}
indicate that our four-dimensional (4D) Universe can be locally
embeded in a five-dimensional (5D) Einstein spacetime.
Liu-Mashhoon-Wesson (LMW) solution \cite{liu95} is a basic solution
in the Space-Time-Matter (STM) theory \cite{stm}, in which a 4D
hypersurface in a 5D Ricci-flat spacetime is chosen as our Universe.

The braneworld scenario has been proposed to describe our Universe
with extra dimension(s) and new physics. In the Randall-Sundrum (RS)
model \cite{rs}, a famous solution that describes the 4D brane in a
5D bulk with a negative cosmological constant was found in
Ref.~\cite{brane}. As an induced matter scenario, the STM model
employs the LMW solution to show its properties. Regardless of the
physical meanings of different solutions, these 5D solutions share
some common features: They describe the embedding of a 4D
submanifold as our Universe to a 5D manifold. The relation between
the STM model and the RS model is shown in Ref.~\cite{equiv}, and
the relation between the STM model and the Dvali-Gabadadze-Porrati
(DGP) model \cite{dgp} is shown in Ref.~\cite{ping}. The relation
between the RS model and the 5D Schwarzshild-AdS black hole has been
analyzed in Refs.~\cite{kraus,global}. It turned out that the LMW
solution is locally isometric to a topological black hole (TBH) in
Ricci-flat spacetime \cite{tbh}. This implies that there should also
exist the cosmological counterpart of the TBH with cosmological
constant.

To generalize the LMW solution, we solve the 5D Einstein equations
with cosmological constant and obtain an exact solution, which
contains two arbitrary functions and three arbitrary constants. By
given the specific forms of the arbitrary functions and constants,
this solution can describe many well-known solutions in a unified
way. We explicitly show the correspondence between this solution and
the one in RS model. This solution is locally isometric to a 5D TBH
with cosmological constant. We find that the metric periodically
depends on the extra dimension if the 5D cosmological constant is
positive. Thus we can compactify the extra dimension on $S^1$ and
then construct a new model, which is different from the RS, DGP, and
STM models, and also distinguishable from the ordinary Kaluza-Klein
cosmology. Many phenomenological issues are worthy further studying.
We also generalize the metric ansatz to the cases with more extra
dimensions and find a linear structure, but the general solution
cannot be obtained.

The paper is organized as follows. In Sec. II we obtain the 5D
solution and propose two cosmological interpretations. In Sec. III
we show that our solution can correspond to the RS model and the
TBH. And we give the interconnections between our solution and other
5D solutions. In Sec. IV we generalize the 5D ansatz to the case
with more extra dimensions and study its features. In the last
section we present the conclusion and discuss some future subjects.

\section{The 5D solution and two cosmological interpretations}
To embed the FRW framework into a 5D spacetime, we use the metric
ansatz
\begin{equation}
ds^2=-B^2(t,y)dt^2+A^2(t,y)\left(\frac{dr^2}{1-kr^2}+r^2d\Omega_2^2\right)+dy^2,\label{eq:ansatz}
\end{equation}
where $k$ is the curvature of the 3D space, and $\Omega_2$ is a 2D
solid angle. This metric is written in the so called Gaussian normal
coordinate system. We solve the Einstein equations with a
cosmological constant $\Lambda$:
\begin{equation}
R_{MN}-\frac{1}{2}g_{MN}R+\Lambda g_{MN}=0,\label{eq:ein}
\end{equation}
where the indices $M$ and $N$ run from 0 to 4. For solving the
Einstein equations, we regard the 5D cosmological constant as an
arbitrary parameter. When we talk about the physical interpretations
of the solution, we may specify that $\Lambda$ is positive or
negative. The $ty$ component of Einstein equations gives
\begin{equation}
B=\frac{\dot{A}}{\mu(t)},\label{eq:ba}
\end{equation}
where a dot denotes a derivative with respect to the time $t$, and
$\mu(t)$ is an arbitrary function of $t$. Then by inserting
Eq.~(\ref{eq:ba}) to the metric ansatz, the $tt$ component of
Einstein equations gives a linear equation of $A^2$,
\begin{equation}
(\partial_y^2+\lambda^2)A^2=2(\mu^2+k),\label{eq:lin}
\end{equation}
where $\lambda=\sqrt{2\Lambda/3}$. Here $\Lambda$ can be both
positive and negative.

Combining Eq.~(\ref{eq:lin}) with other equations, we obtain an
exact solution of Einstein equations. If we require the $\lambda\to
0$ limit to be finite, this general solution is
\begin{eqnarray}
A^2(t,y) &=& \frac{2}{\lambda^2}[\mu^2+k+\lambda v\sin\lambda y\nonumber\\
&&-\sqrt{(\mu^2+k)^2-\lambda^2(\nu^2+K)}\cos\lambda
y],\label{eq:sol1}
\end{eqnarray}
where $\mu\equiv\mu(t)$ and $\nu\equiv\nu(t)$ are arbitrary
functions of $t$, and $K$ is constant. If $\Lambda<0$, by defining
$\lambda=\sqrt{2|\Lambda|/3}$, Eq.~(\ref{eq:sol1}) can be rewritten
as
\begin{eqnarray}
A^2(t,y) &=& \frac{2}{\lambda^2}[-\mu^2-k+\lambda v\sinh\lambda y\nonumber\\
&&+\sqrt{(\mu^2+k)^2+\lambda^2(\nu^2+K)}\cosh\lambda
y].\label{eq:sol2}
\end{eqnarray}
The $\lambda\to 0$ limit of Eq.~(\ref{eq:sol1}) is the LMW solution
\begin{equation}
A^2(t,y)=(\mu^2+k)y^2+2\nu y+\frac{\nu^2+K}{\mu^2+k},\label{eq:lmw}
\end{equation}
which was found by Liu and Mashhoon \cite{stm}, and restudied by
Wesson. Liu has solved the Einstein equations in the 5D bulk and
obtained a solution \cite{liu03}, but that one is apparently
divergent in the $\Lambda\to 0$ limit. The solution (\ref{eq:sol1})
has been essentially obtained in our previous work \cite{sol}.

We should give physical interpretations of our solution. The first
interpretation is the induced matter scenario with etra dimension
unnecessarily compactified. We choose a 4D hypersurface $y=0$, in
which the effective metric is
\begin{equation}
ds^2=-B^2(t,0)dt^2+A^2(t,0)\left(\frac{dr^2}{1-kr^2}+r^2d\Omega_2^2\right).
\end{equation}
By using this 4D metric, the Einstein tensor is
$G_{\mu\nu}=R_{\mu\nu}-\frac{1}{2}g_{\mu\nu}R$. This 4D hypersurface
is not an Einstein spacetime generally, i.e.,
$G_{\mu\nu}+\Lambda_4g_{\mu\nu}\neq 0$, where $\Lambda_4$ is the 4D
cosmological constant. Thus we can define $8\pi G_{\rm
N}T_{\mu\nu}\equiv G_{\mu\nu}+\Lambda_4g_{\mu\nu}$, where $G_{\rm
N}$ is the 4D Newton's constant and $T_{\mu\nu}$ is the
energy-momentum tensor in the $y=0$ hypersurface. This means that
the matter is induced from the extra dimension. Mathematically, the
$\Lambda_4$ and $G_{\rm N}$ are two new parameters. Physically, they
are related to the tension of the brane in the braneworld scenario
\cite{sms}. The Einstein tensor can be calculated as
\begin{eqnarray}
^{(4)}G_0^0 &=& \frac{3(\mu^2+k)}{A^2},\\
^{(4)}G_1^1 &=&
^{(4)}G_2^2=^{(4)}G_3^3=\frac{2\mu\dot{\mu}}{A\dot{A}}+\frac{\mu^2+k}{A^2}.
\end{eqnarray}
where $A$ takes the value in $y=0$ hypersurface. The energy-momentum
tensor is consistent with the perfect fluid. In the following we
will show that this scenario contains the FRW framework with the
scale factor $a(t)$ for specific forms of the functions $\mu(t)$ and
$\nu(t)$.

We first consider the $\Lambda$ cold dark matter ($\Lambda$CDM)
model as a simple case. For the exact solution of the $\Lambda$CDM
model, see Appendix. We solve the scale factor as
\begin{eqnarray}
a(t) &=& a_0\left[\cosh\left(\frac{\sqrt{3\Lambda_4}}{2}(t-t_0)\right)\right.\nonumber\\
&&
\left.+\sqrt{\frac{3}{\Lambda_4}}H_0\sinh\left(\frac{\sqrt{3\Lambda_4}}{2}(t-t_0)\right)\right]^{2/3},\label{eq:lcdm}
\end{eqnarray}
where $\Lambda_4$ is the 4D cosmological constant in our Universe.
For the known function $a(t)$ as the solution of the $\Lambda$CDM
model, if we choose the arbitrary functions $\mu$ and $\nu$ as the
following form:
\begin{eqnarray}
\mu &=& \dot{a}(t),\label{eq:mu}\\
\nu^2 &=& -\frac{1}{6}\Lambda a^4+(\dot{a}^2+k)a^2-K,\label{eq:nu}
\end{eqnarray}
then the $y=0$ hypersurface of Eq.~(\ref{eq:sol1}) is exactly our
Universe described by the $\Lambda$CDM model. Any FRW Universe with
the scale factor $a(t)$ can be reproduced by the $y=0$ hypersurface
in the 5D spacetime, if we choose the arbitrary functions as
Eqs.~(\ref{eq:mu}) and (\ref{eq:nu}). We will show that this
solution mathematically unifies many 5D cosmological solutions in
the next section. We can use the solution (\ref{eq:sol1}) or
(\ref{eq:sol2}) to construct many cosmological models. For example,
we can impose a Z$_2$ symmetry to the spacetime, as in
Ref.~\cite{liko}.

\begin{figure}[]
\includegraphics{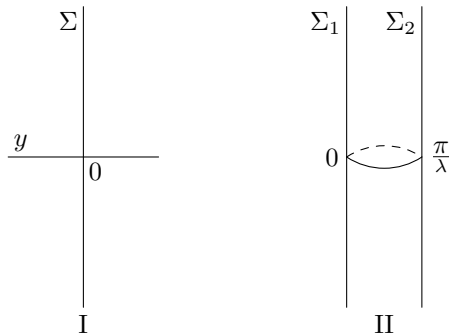}
\caption{\label{fig1} The physical picture of the two
interpretations. In the first one, we take the $y=0$ hypersurface
$\Sigma$ as our Universe. In the second one, the extra dimension is
compactified on S$^1$, but different values of $y$ correspond to
different hypersurfaces.}
\end{figure}

The second interpretation is the induced matter scenario with
compactified extra dimension. We require a positive cosmological
constant in the 5D spacetime. Then the metric periodically depends
on the extra dimension $y$, which implies that the extra dimension
can be compactified on S$^1$. From the metric, we can see that the
scale factor of the extra dimension is constant, which means that
the extra dimension is static. If the Universe begins with a small
volume, the extra dimension is still small now. Different values of
the coordinate $y$ correspond to different 4D spacetimes. We should
take a specific value of $y$ and calculate the 4D effective action
for the Universe. The $y=0$ hypersurface is the same as in the first
interpretation, but the second one is more physical and may have
distinctive signatures. This model is different from the ADD
(Arkani-Hamed, Dimopoulos and Dvali) \cite{add}, RS, DGP, and STM
models, and also different from the ordinary Kaluza-Klein cosmology,
in which the 4D effective metric is independent of the extra
dimension. We will study the phenomenological implications of this
model in our future work. The comparison of the first and the second
interpretations of our solution is shown in Fig.~\ref{fig1}.

\section{Relation to other 5D solutions in a unified way}
\subsection{Relation to the RS, STM, and DGP models}
In the original work of the RS model \cite{rs}, the following
solution of Einstein equations was used to illustrate the main ideas
of the model:
\begin{equation}
ds^2=e^{-2kr_c\phi}\eta_{\mu\nu}dx^\mu dx^\nu+r_c^2d\phi^2,
\end{equation}
where $k$ is around the Planck scale, and the extra dimension $\phi$
is a finite interval whose size is set by $r_c$. A more general
solution with matter in RS model was obtained later \cite{brane}.
The metric ansatz is the same as Eq.~(\ref{eq:ansatz}). The
notations are slightly different in the present work. The
energy-momntum tensor in RS model can be written as
\begin{eqnarray}
\tilde{T}^M_{\quad N} &=& \check{T}^M_{\quad N}|_{\rm
bulk}+T^M_{\quad
N}|_{\rm brane},\\
\check{T}^M_{\quad N}|_{\rm bulk} &=& \textrm{diag}(-\rho_B,-\rho_B,-\rho_B,-\rho_B,-\rho_B),\\
T^M_{\quad N}|_{\rm brane} &=&
\delta(y)\cdot\textrm{diag}(-\rho_b,p_b,p_b,p_b,0).
\end{eqnarray}
where $\rho_B$ is the energy density in the bulk, and $\rho_b$ and
$p_b$ are the energy density and the pressure of the matter in the
brane, respectively. Here $\rho_B$ is a negative constant, and
$\rho_b$ and $p_b$ are functions only of time. The Einstein
equations are $G_{MN}=\kappa^2\tilde{T}_{MN}$. The relation between
$A$ and $B$ is $B=\dot{A}/\dot{a}(t)$, where $a\equiv A(t,0)$. The
solution of $A^2$ is
\begin{eqnarray}
A^2(t,y) &=&
\frac{1}{2}\left(1+\frac{\kappa^2\rho_b^2}{6\rho_B}\right)a^2
+\frac{3C}{\kappa^2\rho_Ba^2}\nonumber\\
&&
+\left[\frac{1}{2}\left(1-\frac{\kappa^2\rho_b^2}{6\rho_B}\right)a^2
-\frac{3C}{\kappa^2\rho_Ba^2}\right]\cosh(\lambda y)\nonumber\\
&&
-\frac{\kappa\rho_b}{\sqrt{-6\rho_B}}a^2\sinh(\lambda|y|),\label{eq:rs}
\end{eqnarray}
where $C$ is constant, and $\lambda=\sqrt{-2\kappa^2\rho_B/3}$. This
$\lambda$ is the same as the one in Eq.~(\ref{eq:sol2}), because the
5D cosmological constant $\Lambda=\kappa^2\rho_B$. The scale factor
$a(t)$ in the brane satisfies the modified Friedmann equation
\begin{equation}
\frac{\dot{a}^2}{a^2}=\frac{\kappa^2}{6}\rho_B+\frac{\kappa^4}{36}\rho_b^2+\frac{C}{a^4}-\frac{k}{a^2}.\label{eq:fried}
\end{equation}

Comparing our solution (\ref{eq:sol2}) with the solution in RS
model, we can see that if we specify some particular form of the
arbitrary functions $\mu$ and $\nu$, Eq.~(\ref{eq:sol2}) can be the
same as Eq.~(\ref{eq:rs}). The correspondence of the functions are
\begin{eqnarray}
\mu^2+k &=&
\left(\frac{\kappa^2}{6}\rho_B+\frac{\kappa^4}{36}\rho_b^2\right)a^2+\frac{C}{a^2},\label{eq:mu}\\
\nu &=& -\frac{\kappa^2}{6}\rho_ba^2,\label{eq:nu}\\
K &=& C.\label{eq:k}
\end{eqnarray}
The integration constant $K$ in Eq.~(\ref{eq:sol2}) is identified
with the integration constant $C$ in Eq.~(\ref{eq:rs}). If we
require $B=1$, then $\mu=\dot{a}$ in the $y=0$ hypersurface. For
this choice of $\mu$, $\nu$, and $K$, Eq.~(\ref{eq:mu}) is the same
as the modified Friedmann equation Eq.~(\ref{eq:fried}). The inverse
transformation is
\begin{eqnarray}
a^2 &=& \frac{2}{\lambda^2}[-\mu^2-k+\sqrt{(\mu^2+k)^2+\lambda^2(\nu^2+K)}],\label{eq:a2}\\
\rho_b &=& -\frac{6\nu}{\kappa^2a^2}.
\end{eqnarray}

The equivalence between the STM model and the RS model has been
demonstrated in Ref.~\cite{equiv}. The 4D effective energy-momentum
tensor \cite{equiv,sms} is the same in both STM and RS models:
\begin{equation}
^{(4)}G_{\mu\nu}=-\Lambda_4q_{\mu\nu}+8\pi G_N\tau_{\mu\nu}+
\kappa_5^4\pi_{\mu\nu}-E_{\mu\nu},\label{eq:sms}
\end{equation}
where $q_{\mu\nu}$ is the induced metric in the brane,
$\tau_{\mu\nu}$ is the energy-momentum tensor in the brane,
$\pi_{\mu\nu}$ is the local quadratic correction, and $E_{\mu\nu}$
is the nonlocal Wyel correction. Here Eq.~(\ref{eq:sms}) is called
Shiromizu-Maeda-Sasaki (SMS) equations \cite{sms}, which are the
effective field equations on the brane (A related work is
Ref.~\cite{aliev}). In the present work, we have explicitly shown
that our solution can be equivalent to the one in RS model in the
$y=0$ hypersurface. The $\Lambda\to 0$ limit of Eq.~(\ref{eq:sol1})
is reduced to the LMW solution in STM model. The correspondence
between the STM model and the DGP model is given in
Ref.~\cite{ping}. Therefore, our solution describes many models in a
unified way mathematically.

\subsection{Relation to the 5D topological black hole}
If we replace the solid angle $\Omega_n$ in the Schwarzschild
solution with an $n$-dimensional Einstein manifold, the metric
remains a solution of Einstein equations. Such a black hole is
called a topological black hole (TBH). The equivalence between the
LMW solution and the TBH in Ricci-flat spacetime has been
demonstrated in Ref.~\cite{tbh}.  We will show that the metric
(\ref{eq:ansatz}) with the solution (\ref{eq:sol1}) can be obtained
by a coordinate transformation of the TBH with cosmological
constant. For the details of this method, see Ref.~\cite{tbh}. The
metric for a 5D TBH is
\begin{equation}
ds_{\rm
TBH}^2=-h(R)dT^2+h^{-1}(R)dR^2+R^2d\Omega_{3(k)}^2,\label{eq:tbh1}
\end{equation}
where $\Omega_{3(k)}$ is a 3D Einstein manifold with curvature $k$.
After the coordinate transformation
\begin{equation}
R=R(t,y),\qquad T=T(t,y),
\end{equation}
Eq.~(\ref{eq:tbh1}) becomes
\begin{eqnarray}
ds^2 &=&
\left(hT_{,t}^2-\frac{R_{,t}^2}{h}\right)dt^2+2\left(hT_{,t}T_{,y}-\frac{R_{,t}R_{,y}}{h}\right)dtdy\nonumber\\
&&+\left(hT_{,y}^2-\frac{R_{,y}^2}{h}\right)dy^2+R^2d\Omega_{3(k)}^2.\label{eq:tbh2}
\end{eqnarray}
The function $R$ can be fixed as $R(t,y)=A(t,y)$. Then by comparing
the coefficients of Eqs.~(\ref{eq:tbh2}) and (\ref{eq:sol1}), the
equations
\begin{eqnarray}
\frac{R_{,t}^2}{\mu^2(t)} &=& h(R)T_{,t}^2-\frac{R_{,t}^2}{h(R)},\label{eq:trans1}\\
0 &=& hT_{,t}T_{,y}-\frac{R_{,t}R_{,y}}{h},\label{eq:trans2}\\
-1 &=& h(R)T_{,y}^2-\frac{R_{,y}^2}{h(R)},\label{eq:trans3}
\end{eqnarray}
are obtained to solve $R$ and $T$. Eqs.~(\ref{eq:trans1}) and
(\ref{eq:trans3}) can be rewritten as
\begin{eqnarray}
T_{,t} &=& \frac{R_{,t}}{h(R)}\sqrt{1+\frac{h(R)}{\mu^2(t)}},\label{eq:tt}\\
T_{,y} &=& \frac{1}{h(R)}\sqrt{R_{,y}^2-h(R)}.\label{eq:ty}
\end{eqnarray}
By substituting these two equations to Eq.~(\ref{eq:trans2}), $R$ is
determined by a single equation
\begin{equation}
R_{,y}^2=h(R)+\mu^2(t).\label{eq:rty}
\end{equation}
After $R(t,y)$ is obtained, we can substitute it to
Eqs.~(\ref{eq:tt}) and (\ref{eq:ty}) to solve $T(t,y)$. The
integrable condition is satisfied for these equations.

Starting with Eq.~(\ref{eq:tbh1}) and solving the Einstein equation
with a cosmological constant $\Lambda$, we can obtain
\begin{equation}
h(R)=k-\frac{K}{R^2}-\frac{1}{6}\Lambda R^2.
\end{equation}
We solve Eq.~(\ref{eq:rty}) and find
\begin{equation}
R^2=\frac{2}{\lambda^2}[\mu^2+k+\sqrt{(\mu^2+k)^2-\lambda^2K}\sin(\lambda
y+\varphi)],\label{eq:solr}
\end{equation}
where $\varphi\equiv\varphi(t)$ is an arbitrary function from the
integration. This is consistent with Eq.~(\ref{eq:sol1}). By
applying the identity
$a\sin\theta+b\cos\theta=\sqrt{a^2+b^2}\sin(\theta+\varphi)$, where
$\tan\varphi=b/a$ to Eq.~(\ref{eq:sol1}), and comparing the result
with Eq.~(\ref{eq:solr}), we obtain
\begin{equation}
\tan\varphi=-\frac{1}{\lambda\nu}\sqrt{(\mu^2+k)^2-\lambda^2(\nu^2+K)}.
\end{equation}
With the help of the equivalence between our solution and the TBH,
we directly obtain the Kretschmann scalar to be
\begin{equation}
I=R_{MNPQ}R^{MNPQ}=\frac{72K^2}{A^8}+\frac{10\Lambda^2}{9}
\end{equation}
For the LMW solution, the Kretschmann scalar is
$I=R_{MNPQ}R^{MNPQ}=72K^2/A^8$.

In the 4D case, if we attempt to transform the 4D Schwarzshild
solution in the same way, we need to solve the equation
\begin{equation}
\frac{\partial R}{\partial y}=\sqrt{1-\frac{K}{R}+\mu^2(t)}.
\end{equation}
Although the integration can be evaluated out, we cannot solve $R$
explicitly. In the 6D case, the integration cannot be evaluated out.
This implies that the 5D metric (\ref{eq:ansatz}) is special, e.g.,
if we modify the $\Omega_2$ to $\Omega_3$, we cannot have an
explicit solution as the counterpart of the 6D TBH. The specialty of
5D solution has been noticed by Seahra and Wesson \cite{tbh}, but
they did not consider the case when the cosmological constant
exists. Luckily, Eq.~(\ref{eq:rty}) can also be explicitly solved in
the more general case with cosmological constant.

In addition, it has shown that the following Fukul-Seahra-Wesson
(FSW) solution \cite{fsw} is equivalent to LMW solution after Wick
rotations:
\begin{equation}
ds_{\rm
FSW}^2=-d\tau^2+b^2(\tau,w)d\Omega_{3(k)}^2+\frac{b_{,w}^2(\tau,w)}{\zeta^2(w)}dw^2,
\end{equation}
where
\begin{equation}
b^2(\tau,w)=[\zeta^2(w)-k]\tau^2+2\chi(w)+\frac{\chi^2(w)-K}{\zeta^2(w)-k}.
\end{equation}
Similarly, we can easily generalize the FSW solution to the case
with cosmological constant from Eq.~(\ref{eq:sol1}).

\subsection{Relations to other solutions}
The following Lema\^{\i}tre-Tolman-Bondi (LTB) metric \cite{ltb} has
been used to describe the inhomogeneous Univese:
\begin{equation}
ds^2=-dt^2+\frac{A'(r,t)^2}{1+f(r)}dr^2+A(r,t)^2d\Omega_2^3,\label{eq:ltb4}
\end{equation}
where the unknown functions $A(t,r)$ and $f(r)$ are to be solved
with the energy-momentum tensor. The metric (\ref{eq:ansatz}) is
essentially a 5D LTB metric
\begin{equation}
ds^2=-dt^2+\frac{A'(r,t)^2}{1+f(r)}dr^2+A(r,t)^2d\Omega_{3(k)}^3,\label{eq:ltb5}
\end{equation}
which can be transformed to the same form as Eq.~(\ref{eq:ansatz})
by a double Wick rotation
\begin{equation}
y\to it,\qquad t\to ir,\qquad r\to\tilde{r}.
\end{equation}
This implies that the 5D LTB metric can be explicitly solved in
Einstein spacetime. However, the 4D LTB metric cannot be explicitly
solved (it can be solved in parametric form). It is a coincidence
that $A^2$ satisfies a linear equation in 5D case, while $A^2$
satisfies a nonlinear equation in 4D or other cases generally. This
confirms that the 5D TBH and the 5D LTB metric are special. In fact,
the linear structure for the ansatz (\ref{eq:ansatz}) will be more
clear in the case of more extra dimensions, as the next section
shows.

The metric ansatz can be written as the Weyl type
\begin{equation}
ds^2=e^{2\beta(t,y)}(-dt^2+dy^2)+e^{2\alpha(t,y)}g_{ij}dx^idx^j,\label{eq:gen}
\end{equation}
which was used in Ref.~\cite{weyl}. For the generalized Weyl
solution \cite{gws}, see Appendix. The solution for
Eq.~(\ref{eq:gen}) in Ricci-flat spacetime can be directly obtained
as a generalized Weyl solution. However, the generalized Weyl
solution is only applicable to the vacuum Einstein equations. We
must simplify the ansatz (\ref{eq:gen}) to solve Einstein equations
with cosmological constant. To make the ansatz (\ref{eq:ansatz}) as
general as Eq.~(\ref{eq:gen}), we should add another unknown
function before $dy^2$. The physical meaning of the simplification
as Eq.~(\ref{eq:ansatz}) is setting the extra dimension to be
static. Mathematically, this simplification enhances the symmetry of
the spacetime. The metric (\ref{eq:gen}) contains three Killing
vectors, while Eq.~(\ref{eq:ansatz}) contains four Killing vectors
essentially, because it can be transformed to a 5D TBH.

The relations between the solutions are as follows:
\begin{displaymath}
\xymatrixcolsep{3pc}\xymatrixrowsep{3pc}\xymatrix{
& {\rm LTB} \ar@{<->}[d]^{\cal WR} & \\
{\rm RS} & {\rm RMZ} \ar[d]^{\Lambda\to 0} \ar[l]_{\Lambda<0,\textrm{ Z}_2} \ar@{<->}[r]^{\cal CT} & {\rm TBH}\\
& {\rm LMW} \ar[ul]^{\textrm{Z}_2} \ar@{<->}[ur]_{\cal CT} &}
\end{displaymath}
Here $\mathcal{WR}$ denotes a double Wick rotation, and
$\mathcal{CT}$ denotes a coordinate transformation. The related
solutions are as follows:
\begin{itemize}
\item LTB denotes the 5D LTB metric, Eq.~(\ref{eq:ltb5}).
\item RS denotes a solution in the RS model, Eq.~(\ref{eq:rs}),
which was found by Bin\'{e}truy, Deffayet, Ellwanger, and Langlois.
\item RMZ denotes our solution, Eq.~(\ref{eq:sol1}).
\item TBH denotes the topological black hole, Eq.~(\ref{eq:tbh1}).
\item LMW denotes the LMW solution, Eq.~(\ref{eq:lmw}).
\end{itemize}
The generality of our solution originates from two arbitrary
functions and three arbitrary constants that it contains. An
important problem is whether we can obtain some new physics from
this solution. We can look the relation between these solutions in
the physical point of view. The braneworld scenarios describe a
brane in a 5D bulk with a cosmological constant $\Lambda$. If
$\Lambda<0$, we have the RS model, and if $\Lambda=0$, we have the
DGP model. Both of these two models have rich phenomenological
issues. Consequently, there seems a vacancy in the $\Lambda>0$ case.
The second interpretation of our solution may correspond to a new
model to fill in this vacancy, as shown in Table \ref{tab:t1}.

\begin{table}[h]
\caption{\label{tab:t1} Three cases of the 5D cosmological constant}
\begin{ruledtabular}
\begin{tabular}{cc}
Cosmological constant & Physical model\\
\hline
$\Lambda<0$ & Randall-Sundrum model (Type II)\\
$\Lambda=0$ & Dvali-Gabadadze-Porrati model\\
$\Lambda>0$ & The second interpretation
\end{tabular}
\end{ruledtabular}
\end{table}

\section{More extra dimensions}
We want to generalize the codimension one scenario to the case with
more extra dimensions. The codimension two brane has been studied,
such as in Ref.~\cite{wang}. If the number of extra dimensions is
$n$, we propose a metric ansatz as
\begin{eqnarray}
ds^2 &=& -B^2(t,Y)dt^2+A^2(t,Y)\left(\frac{dr^2}{1-kr^2}+r^2d\Omega_2^2\right)\nonumber\\
&& +\sum_idy_i^2,
\end{eqnarray}
where $y_1,y_2,...y_n$ are coordinates of the extra dimensions, and
$Y\equiv(y_1,y_2,\cdots,y_n)$. Define $\lambda=\sqrt{2\Lambda/3}$,
where $\Lambda$ is the $(4+n)$-dimensional cosmological constant.
The relation between $A$ and $B$ is also $B=\dot{A}/\mu(t)$. We find
that as along as the $A^2$ satisfies the following Helmholtz
equation with source term:
\begin{equation}
(\sum_i\partial_i^2+\lambda^2)A^2=2(\mu^2+k),\label{eq:lap}
\end{equation}
the tensor $G_{\mu\nu}+\Lambda g_{\mu\nu}$ will be as the form
\begin{displaymath}
\left(
\begin{array}{cc}
0_{4\times 4} & 0\\
0 & X_{n\times n}
\end{array}
\right).
\end{displaymath}
The linear equation (\ref{eq:lap}) guarantees that the 4D part of
the Einstein equations is satisfied. But the other parts of Einstein
equations are nonlinear and cannot be solved generally. For example,
in the Ricci-flat spacetime, other parts of Einstein equations can
be written as
\begin{equation}
[(A^2)_{,ij}A^6]_{,t}=0,
\end{equation}
for any $i,j=0,1,\cdots,n$. We can add some other fields in extra
dimension to compensate the non-zero components, and other fields
may stabilize the extra dimensions.

We can also consider the case that both $A$ and $B$ are independent
of time. The 5D ansatz is
\begin{equation}
ds^2=-B^2(y)dt^2+A^2(y)\left(\frac{dr^2}{1-kr^2}+r^2d\Omega_2^2\right)+dy^2,\label{eq:indept}
\end{equation}
which cannot be directly treated as the special case of the ansatz
(\ref{eq:ansatz}), because $B=0$ if $\dot{A}=0$. However, if we
require $B=1$ and then $\mu=\dot{A}$, we have $\mu=0$ when
$\dot{A}=0$. We can directly check that the following $A$ with $B=1$
is an exact solution:
\begin{equation}
A^2(y)=\frac{2}{\lambda^2}[k+\lambda\nu\sin\lambda
y-\sqrt{k^2-\lambda^2(\nu^2+K)}\cos\lambda y],\label{eq:ay}
\end{equation}
where $\nu$ is constant. The $\lambda\to 0$ limit is
\begin{equation}
A^2(y)=ky^2+2\nu y+\frac{\nu^2+K}{k},
\end{equation}
as the LMW solution implies. Here Eq.~(\ref{eq:ay}) is a special
solution for the ansatz (\ref{eq:indept}). In the Ricci-flat
spacetime, the following solution:
\begin{equation}
A^2(y)=ky^2+c,\qquad B^2(y)=\frac{y^2}{ky^2+c},
\end{equation}
where $c$ is constant, is also a special solution for the ansatz
(\ref{eq:indept}).

The higher-dimensional ansatz is
\begin{equation}
ds^2=
-B^2(Y)dt^2+A^2(Y)\left(\frac{dr^2}{1-kr^2}+r^2d\Omega_2^2\right)+\sum_idy_i^2.
\end{equation}
The function $A$ also satisfies a linear equation
\begin{equation}
(\sum_i\partial_i^2+\lambda^2)A^2=2k,
\end{equation}
where $\lambda$ is defined as above. As long as $A$ satisfies this
equation, the $tt$ component of Einstein equations will be
satisfied. However, other nonlinear equations of $A$ and $B$ cannot
be written as a simple form. We can show another special solution.
In the Ricci-flat spacetime, the 6D ansatz
\begin{eqnarray}
ds^2=-B^2dt^2+A^2(dr^2+r^2d\Omega_2^2)+dy^2+dz^2,
\end{eqnarray}
gives a Weyl type solution
\begin{equation}
A^2(y,z)=y^2-z^2,\qquad B^2(y,z)=\frac{y^2z^2}{y^2-z^2}.
\end{equation}
The analysis of this solution is beyond the scope of this paper.

\section{Conclusion and discussion}
We have obtained an exact solution of 5D Einstein equations with
cosmological constant and shown the interconnections between this
solution and other solutions. Two interpretations to this solution
are given. In the first interpretation, we take the $y=0$
hypersurface as our Universe with induced matter. We have
demonstrated the mathematical equivalence of our solution and the
solution in RS model and the TBH. In the second interpretation, we
require the 5D cosmological constant to be positive and thus
compactify the extra dimension on S$^1$. This scenario is similar to
the Kaluza-Klein cosmology, but the 4D effective metric depends on
the coordinate of the extra dimension. We also propose a metric
ansatz with more extra dimensions, and find that the 4D part of
Einstein equations is reduced to a Helmholtz equation with source
term. This linear structure in the 5D case shows that the 5D TBH and
the 5D LTB matric are special. We also give some special solutions
in the time-independent case.

We shall discuss some possible future developments of our work. The
phenomenological implications of our solution, especially the
$\Lambda>0$ case, should be studied in details.
\begin{itemize}
\item Different coordinate systems may cover different patches of the whole manifold.
In the Penrose diagram of the extended Schwarzshild-(A)dS manifold,
which patch does this solution cover?
\item If the 5D cosmological constant is positive,
is it possible to construct a physical model, which is parallel to
the RS and DGP model?
\item The corrections to the observable quantities
should be calculated. What is the 4D effective action after the
dimension reduction? What is the correction to the Newtonian
potential in 4D?
\item The currently accelerating expansion of our
Universe may be due to modified gravity \cite{meng} or
higher-dimensional effects. Can this solution give some new insight
to the 5D or 4D gravitational physics?
\item The stabilization of the extra dimension(s). The moduli
stabilization may be related to dark energy \cite{brg}. Thus, what
will happen in our case?
\item Can this solution explain inflation and large scale structure
of our Universe?
\end{itemize}

\section*{ACKNOWLEDGMENTS}
We thank the KITPC for hospitality during the completion of this
work. J.R. thanks Peng Wang, Prof. S.-H. Henry Tye and Prof. Jiro
Soda for helpful discussions on the compactification of the extra
dimension. X.H.M. was supported by NSFC under No. 10675062, and BK21
Foundation. L.Z. was supported by NSFC under No. 90403014. This
research was supported in part by the Project of Knowledge
Innovation Program (PKIP) of Chinese Academy of Sciences.

\appendix
\section{Mathematical notes}
The solution of the $\Lambda$CDM model is presented as follows. The
FRW metric is
\begin{equation}
ds^2=-dt^2+a^2(t)(dr^2+d\Omega_2^2),
\end{equation}
where $a(t)$ is the scale factor to be solved. The content of the
Universe is assumed to be a perfect fluid, whose energy-momentum
tensor is $T^{\mu}_{\;\;\nu}=\textrm{diag}(-\rho,p,p,p)$. Einstein
equations with a cosmological constant $\Lambda$ are reduced to
Fiedmann equations,
\begin{eqnarray}
\frac{\dot{a}^2}{a^2} &=& \frac{8\pi
G}{3}\rho+\frac{\Lambda}{3},\nonumber\\
\frac{\ddot{a}}{a} &=& -\frac{4\pi G}{3}(\rho+3p)+\frac{\Lambda}{3}.
\end{eqnarray}
Note that this $\Lambda$ is $\Lambda_4$ in Sec. II. For the
$\Lambda$CDM model, the equation of state is $p=0$. The solution of
$a(t)$ is Eq.~(\ref{eq:lcdm}). The $H$-$a$ relation can be solved as
\begin{equation}
H(a)=H_0^2\left[\left(1-\frac{\Lambda}{3H_0^2}\right)\left(\frac{a_0}{a}\right)^3+\frac{\Lambda}{3H_0^2}\right],
\end{equation}
where $H_0$ is the Hubble constant, $\Omega_m$ is the ratio of the
matter density. The relation between $\Omega_m$ and $\Lambda$ is
$\Omega_m=1-\Lambda/(3H_0^2)$. The linearization of friedmann
equations was studied in Ref.~\cite{ren1}, and the physical analysis
of some special cases was studied in Ref.~\cite{ren2}.

The generalized Weyl solution \cite{gws} is a successful application
of the integrable theory to general relativity. It is valid for
vacuum Einstein equations with a $D$-dimensional metric that admits
$D-2$ orthogonal Killing vector fields. The metric ansztz is
\begin{equation}
ds^2=\sum_{i=1}^{D-2}\epsilon_ie^{2U_i}(dx^i)^2+e^{2C}dZd\bar{Z},
\end{equation}
where $U_i$ and $C$ are functions of $Z$ and $\bar{Z}$ only, and
$\epsilon_i=\pm 1$. The functions $U_i$ satisfy a Laplace equation
\begin{equation}
\partial_Z\partial_{\bar Z}\exp(\sum_jU_j)=0.
\end{equation}
Thus the solution can be written as
$\sum_jU_j=\log(w(Z)+\tilde{w}(\bar{Z}))$, and the function $C$ is
given by $C=\frac{1}{2}\log(\partial_Zw\partial_{\bar
Z}\tilde{w})+\nu$, where $\nu$ is determined by
\begin{eqnarray}
\partial_Z\nu &=& -\frac{w+\tilde{w}}{\partial_Zw}\sum_{i<j}\partial_ZU_i\partial_ZU_j,\\
\partial_{\bar{Z}}\nu &=& -\frac{w+\tilde{w}}{\partial_{\bar Z}\tilde{w}}\sum_{i<j}\partial_{\bar Z}U_i\partial_{\bar
Z}U_j.
\end{eqnarray}
The integrable condition for $\nu$ has been satisfied.

\end{document}